# Biphoton focusing for two-photon excitation

Magued B. Nasr, Ayman F. Abouraddy, Mark C. Booth, Bahaa E. A. Saleh[†], Alexander V. Sergienko, and Malvin C. Teich

*Quantum Imaging Laboratory[‡], Departments of Electrical & Computer Engineering and Physics, Boston University, 8 Saint Mary's Street, Boston, MA 02215, USA*

Michael Kempe and Ralf Wolleschensky

*Carl Zeiss Jena GmbH, Carl Zeiss Group, Carl-Zeiss-Promenade 10, 07745 Jena, Germany*

**Abstract**

We study two-photon excitation using biphotons generated via the process of spontaneous parametric down-conversion in a nonlinear crystal. We show that the focusing of these biphotons yields an excitation distribution that is essentially the same as the distribution of one-photon excitation at the pump wavelength. We also demonstrate that biphoton excitation in the image region yields a distribution whose axial width is approximately that of the crystal thickness and whose transverse width is that of the pump at the input to the crystal.

---

[†] Electronic address: besaleh@bu.edu
[‡] URL: http://www.bu.edu/qil

# I. INTRODUCTION

Two-photon excitation is a nonlinear optical process in which the simultaneous absorption of two photons excites an atomic system to a higher energy state. This process, first studied by Göppert-Mayer in 1931 [1], has found use in various applications including microscopy [2,3], spectroscopy [4], and lithography [5]. Since the arrival times and locations of photons generated by traditional light sources are random, an appreciable rate of simultaneous two-photon absorption can only be achieved with an intense light source that is tightly focused in space and time [3].

Some quantum sources of light exhibit statistical behavior that is more suited to such a process. One such source is provided by the process of spontaneous parametric down-conversion (SPDC). In this process an intense optical beam (pump) illuminates a nonlinear crystal (NLC). Some of the pump photons disintegrate into photon pairs (biphotons), traditionally called signal and idler, that conserve the energy and momentum of the parent pump photon [6-8]. The signal and idler photons are created almost simultaneously in the NLC [9]. It is expected that the use of this source for the purpose of two-photon excitation will yield a two-photon absorption rate comparable with that produced by a random source of light, but at a far lower photon flux [10]. It has also been suggested that such a source could find use in two-photon lithography [11].

In this paper we study the distribution of two-photon excitation using biphotons, which we refer to as *biphoton excitation*. We begin by reviewing one- and two-photon excitation distributions using classical light (Sec. II). Using the formalism of biphoton excitation (Sec. III), we study two simple cases: biphoton excitation in the focal region (Sec. IV); and biphoton excitation in the image region (Sec. V). We determine the effect of the various physical parameters of the SPDC process and the optical delivery system on the biphoton excitation distribution. All lenses used are assumed to be thin and, for simplicity without loss of generality, we assume a one-dimensional (in the transverse direction) optical system.

## II. REVIEW OF ONE- AND TWO-PHOTON EXCITATION IN THE FOCAL REGION

For a monochromatic plane wave of wavelength $\lambda_p$ focused by a thin lens of focal length $f$ and aperture $D$, the rate of *one-photon excitation*, $G^{(1)}(x, z)$, which is proportional to the intensity in the focal region, is given by

$$G^{(1)}(x, z) \propto \frac{1}{U} A(X, Z),$$ (2.1)

where $U = 1 + \dfrac{z}{f}$, $X = \dfrac{2x/x_c}{1 + \dfrac{z}{f}}$, $Z = \dfrac{2z/z_c}{1 + \dfrac{z}{f}}$, and

$$A(X, Z) = \left| \int_{-0.5}^{0.5} d\beta \exp(-j2\pi X\beta) \exp(-j\pi Z\beta^2) \right|^2.$$ (2.2)

The variables $z_c = 2\lambda_p F_\#^2$ and $x_c = 2\lambda_p F_\#$ are the characteristic axial and transverse distances, respectively, and $F_\# = \dfrac{f}{D}$ is the *F*-number of the lens.

We now consider *two-photon excitation*. A monochromatic plane wave of wavelength $\lambda_o = 2\lambda_p$ is focused onto the focal region, as illustrated in Fig. 1(a). The rate of two-photon excitation, i.e., the rate at which photons are absorbed in pairs, is proportional to the square of the incident intensity [12], $G^{(2)}(x,z) \propto |E(x,z)|^4$, where $E(x,z)$ is the incident field. Therefore,

$$G^{(2)}(x,z) \propto \frac{1}{U^2} A^2\left(\frac{X}{2}, \frac{Z}{2}\right). \tag{2.3}$$

The axial distribution at $x = 0$ is

$$G^{(2)}(0,z) \propto \frac{1}{U^2} A^2\left(0, \frac{Z}{2}\right), \tag{2.3}$$

and the transverse distribution at the focal plane ($z = 0$) can be shown to be

$$G^{(2)}(x,0) \propto \operatorname{sinc}^4\left(\frac{X}{2}\right). \tag{2.4}$$

Typically $z_c \ll f$ and $U \approx 1$, so that $X \approx \frac{2x}{x_c}$ and $Z \approx \frac{2z}{z_c}$. In this case the distributions of one-photon excitation $G^{(1)}(x,z)$ and two-photon excitation $G^{(2)}(x,z)$ are independent of the focal length $f$. The axial sections of $G^{(1)}(x,z)$ (dotted), and $G^{(2)}(x,z)$ (solid), are plotted in Fig. 2(a). The abscissa is normalized through division by $z_c$. The transverse sections are plotted in Fig. 2(b) and the abscissa is normalized through division by $x_c$.

From Figs. 2(a) and 2(b) we deduce that the axial width (FWHM) of one-photon excitation is $3.5 z_c$, and the transverse width is $0.44 x_c$. The axial width of two-photon excitation is $5.0 z_c$, and the transverse width is $0.64 x_c$.

# III. BIPHOTON EXCITATION

To study the axial and transverse distributions of biphoton excitation we present a brief review of the results reported in Ref. [13].

Consider SPDC from a NLC of thickness $\ell$ pumped by a normally incident monochromatic beam of angular frequency $\omega_p = \dfrac{2\pi c}{\lambda_p}$ and transverse field distribution $E_p(x)$. The signal $s$, and idler $i$, beams are both transmitted through a linear optical system of impulse response function $h(x, z; x'; \omega)$, examples of which are shown in Figs. 1(b) and 3. In this paper we consider crystals of thickness such that $\ell >> \lambda_p$, which is the practical case.

The biphoton excitation, which is proportional to the two-photon coincidence at the same space-time point, is

$$G_b^{(2)}(x, z) = \left| \iiint dq_s dq_i d\omega_s \, \Lambda(q_s, q_i, \omega_s) H(x, z; q_s; \omega_s) \, H(x, z; q_i; \omega_p - \omega_s) \right|^2 , \qquad (3.1)$$

where the transfer function $H(x, z; q; \omega) = \int dx' \, h(x, z; x'; \omega) e^{iqx'}$. It is understood that all integration limits extend from $-\infty$ to $\infty$ unless otherwise indicated.

The function $\Lambda$ in Eq. (3.1) is related to the pump and NLC parameters by

$$\Lambda(q_s, q_i, \omega_s) = \widetilde{E}_p(q_s + q_i)\widetilde{\zeta}(q_s, q_i; \omega_s), \qquad (3.2)$$

where

$$\widetilde{\zeta}(q_s, q_i; \omega_s) = \ell \, \mathrm{sinc}\left( \frac{\ell}{2\pi} \Delta r \right) \exp\left( -j \frac{\ell}{2} \Delta r \right), \qquad (3.3)$$

with $\widetilde{E}_p(q) = \int dx E_p(x) e^{-iqx}$ , $\Delta r = r_p - r_s - r_i$ , $r_j = \sqrt{n_j^2(\omega)\dfrac{\omega^2}{c^2} - q_j^2}$ , and $j = p,s,i$ ; $r_j$ ,

$q_j$ are the longitudinal and transverse components of the momentum vector, respectively, and $n_j(\omega)$ is the index of refraction of the NLC at $\omega$ . The NLC is birefringent and thus its index of refraction may be ordinary or extraordinary, depending on the SPDC configuration: type-I or type-II [7,8]. In this paper we study type-I SPDC where the polarization of the pump is extraordinary while those for the signal and idler are ordinary. We consider $\beta$-barium borate (BBO) crystals for our computations.

The function $\Lambda(q_s, q_i, \omega_s)$ defined by Eq. (3.2) confines the spatial and spectral frequencies of the down-converted pairs to certain bands. We differentiate here between two cases: the collinear case, where the majority of SPDC biphotons are emitted in directions centered about the direction of the pump, and the non-collinear case, where the majority of SPDC is centered about some off-axis directions. The crystal cut-angle, $\theta_{cut}$ , determines which case is produced and changing $\theta_{cut}$ will change the output gradually from one case to the other [14].

## IV. BIPHOTON EXCITATION IN THE FOCAL REGION

Consider the configuration illustrated in Fig. 1(b) where collinear biphotons generated by SPDC are focused onto a thick target. A plane-wave pump is assumed and a narrow spectral filter is used to limit the signal and idler (biphoton) bandwidths, to a narrow range about the degenerate wavelength $\lambda_o = 2\lambda_p$. The transfer function for this system is [8,15]

$$H(x,z;q;\omega) = \frac{1}{\sqrt{\lambda f U}} \exp(j\frac{\pi x^2}{\lambda f U}) \exp(-j\frac{\lambda d}{4\pi}q^2) \widetilde{P}_g\left(q - \frac{2\pi x}{\lambda f U}\right), \qquad (4.1)$$

where $\widetilde{P}_g(q)$ is the Fourier transform of the generalized pupil function

$P_g(x) = p(x)\exp\left(-j\frac{\pi z}{\lambda U f^2}x^2\right)$, $p(x)$ is the pupil function which is taken to be

rectangular of width $D$, and $d$ is the distance between the NLC and lens. Using Eqs. (3.1)-(3.3) and (4.1), one may show that

$$G_b^{(2)}(x,z) \propto \left| \int_{-0.5}^{0.5}\int_{-0.5}^{0.5} d\alpha d\beta \exp\left(-j\pi\frac{Z}{2}(\alpha^2 + \beta^2)\right)\exp\left(-j\pi X(\alpha + \beta)\right)g(\alpha - \beta) \right|^2, \qquad (4.2)$$

with

$$g(\alpha - \beta) = \int d\rho \, \mathrm{sinc}\left(\mu_1\rho^2\right)\exp\left(-j2\pi\mu_2\rho^2\right)\exp\left(j2\pi(\alpha - \beta)\rho\right) \; ; \qquad (4.3)$$

$\mu_1 = \frac{\ell_{eq}}{f}\frac{1}{N_f}$, $\mu_2 = \frac{\frac{\ell_{eq}}{2}+d}{f}\frac{1}{N_f}$, $N_f = \frac{D^2}{\lambda_o f}$ is a Fresnel number, $\ell_{eq} = \frac{\ell}{n_o(\omega_o)}$ is an

equivalent crystal thickness, and $n_o(\omega_o)$ is the ordinary refractive index of the NLC at the degenerate angular frequency $\omega_o = \frac{2\pi c}{\lambda_o}$. Since $\mu_1 \ll 1$ and $\mu_2 \ll 1$ for all practical values of $\ell$ and distances $d$, Eq. (4.3) simplifies to $g(\alpha - \beta) = \delta(\alpha - \beta)$. As a result Eq. (4.2) gives

$$G_b^{(2)}(x,z) \propto \frac{1}{U^2}A(X,Z), \qquad (4.4)$$

and therefore the transverse biphoton excitation is

$$G_b^{(2)}(x,0) \propto \text{sinc}^2(X) \ .$$  (4.5)

Except for the $\dfrac{1}{U^2}$ factor in Eq. (4.4), which is approximately unity for all practical cases, the biphoton excitation $G_b^{(2)}(x,z)$ is identical to the one-photon excitation at the pump wavelength, $G^{(1)}(x,z)$, given in Eq. (2.1) and illustrated in section in Figs. 2(a) and 2(b).

# V. BIPHOTON EXCITATION IN THE IMAGE REGION

We now consider the delivery of biphotons to a thick target by a thin lens in an imaging configuration of magnification $M$. We first consider an ideal system, and compare it to systems including non-ideal effects taken one at a time.

## A. Ideal imaging system

We consider an ideal lens with no aperture; assume unity magnification ($M$=1) and a spectral filter of narrow bandwidth centered about $\lambda_o$ (see Fig. 3). For points in the vicinity of the image plane ($z = 0$), the transfer function of this system is [8,14],

$$H(x,z;q;\omega) \propto \exp\left(-j\frac{\lambda z}{4\pi U}q^2\right)\exp\left(-j\frac{xq}{U}\right)\exp\left(j\frac{\pi x^2}{2\lambda f}\left(1+\frac{1}{U}\right)\right).$$  (5.1)

For a plane-wave pump it can be shown that the distribution of biphoton excitation is

$$G_{bi}^{(2)}(x,z) \propto \left|\int dq\, \widetilde{\zeta}(q,-q;\omega_o)\exp\left(-j\frac{\lambda_p z}{\pi U}q^2\right)\right|^2 ,$$  (5.2)

where the subscript $bi$ denotes this ideal biphoton system. Since the function $G_{bi}^{(2)}(x,z)$ is independent of $x$, we denote it hereafter as $G_{bi}^{(2)}(z)$.

For collinear SPDC, plots of $G_{bi}^{(2)}(z)$ are shown in Fig. 4 for several values of $\ell$, assuming that $f >> \ell_{eq}$ (i.e., $U \approx 1$). It is evident that the three distributions have the same general shape but are scaled in width and height as $\ell$ varies. Using a pump of wavelength $\lambda_p = 532$ nm, the collinear cut-angle is $\theta_{cut} = 22.88º$. As the cut-angle is changed from the collinear value, the distribution gradually narrows and eventually becomes bounded by a rectangular function of width $\ell_{eq}$. This effect is more pronounced for thicker crystals. Figure 5 illustrates $G_{bi}^{(2)}(z)$ for several crystal cut-angles corresponding to collinear and non-collinear SPDC in a 2-mm thick crystal. The abscissa is normalized through division by $\ell_{eq}$. The height increases to a peak value and then decreases as the transition from collinear to non-collinear occurs. This effect is shown in Fig. 6.

We have also examined the case of very short focal length, when the condition $f >> \ell_{eq}$ is not satisfied. In this case the axial distribution is narrowed. For example, for the case of collinear SPDC with $\dfrac{f}{\ell_{eq}} = 1$ in a 2-mm thick crystal, the axial distribution narrows by 30%. In the remainder of this paper we retain the assumption $f >> \ell_{eq}$.

**B. Effect of finite lens aperture**

The effect of the lens aperture is governed by the ratio between the angle subtended by the lens, $\theta_{lens} = \dfrac{M}{2(M+1)F_{\#}}$, and the angle of the cone of SPDC emission, $\theta_{SPDC}$ (see Fig. 1b). Using the NLC function $\tilde{\zeta}(q_s, q_i; \omega_s)$, $\theta_{SPDC}$ is estimated to be $\theta_{SPDC} = \sqrt{2n_o(\omega_o)\Delta n + \dfrac{\lambda_O}{\ell_{eq}}}$. The refractive index difference $\Delta n = n_o(\omega_o) - n_e(\omega_p, \theta_{cut})$ is positive for the non-collinear case and is zero for the collinear case. A negative value for $\Delta n$ indicates non-degenerate emission, which we do not consider here. The parameter $n_e(\omega_p, \theta_{cut})$ is the extraordinary index of refraction of the pump. We have computed the biphoton axial distribution, $G_b^{(2)}(0, z)$, for several values of the ratio $\delta = \dfrac{\theta_{SPDC}}{\theta_{lens}}$ and the results are shown in Fig. 7. The finite lens aperture has the effect of widening $G_b^{(2)}(0, z)$. However, if $\delta < 0.65$, we can neglect the effect of the finite aperture.

### C. Effect of finite transverse width of the pump

For a pump confined to an aperture of characteristic width $b$ such that $b >> \sqrt{\ell_{eq}\lambda_p}$, which is the case we consider here, the shape of $G_b^{(2)}(0, z)$ is identical to that of the ideal system $G_{bi}^{(2)}(z)$. If the limit $b >> \sqrt{\ell_{eq}\lambda_p}$ is not satisfied, the diffraction within the crystal has to be treated rigorously. By substituting $H(x, z = 0; q; \omega)$ [Eq. (5.1)] in Eq. (3.1), it can be shown that the transverse distribution at the in-focus imaging plane ($z = 0$) is the image of the pump, just as predicted by geometrical optics.

## D. Effect of SPDC bandwidth

Finally, we study the effect of increasing the biphoton bandwidth on $G_b^{(2)}(0,z)$. Substituting the transfer function $H(x=0,z;q;\omega)$ [Eq. (5.1)] into Eq. (3.1), we have computed $G_b^{(2)}(0,z)$ assuming a lens with no chromatic aberration. The results are presented in Figs. 8 and 9. From these plots of $G_b^{(2)}(0,z)$ it is clear that the height increases and the width decreases as the biphoton bandwidth is increased. The effect is more pronounced for thicker crystals and for larger cut-angles.

The change in the axial distribution arising from increasing biphoton bandwidth is shown in Figs. 8(a) and 8(b) for crystals with $\ell = 10$ mm and $\ell = 2$ mm, respectively. The plots in these two figures are normalized to have unity peak values in order to easily observe the change in the width. The relative dependence of the peak value of $G_b^{(2)}(0,z)$ on the biphoton bandwidth is shown in Fig. 9 for both crystal thicknesses. The biphoton bandwidth is centered about the degenerate angular frequency $\omega_o = \dfrac{2\pi c}{\lambda_o}$, where $\lambda_o = 2\lambda_p = 1064$ nm is the degenerate wavelength used in the computations. For each plot in Fig. 8 the bandwidth value (BW) is normalized to $\omega_o$ and is also given in nm. In computing Figs. 8 and 9, we have chosen $\theta_{cut}$ to correspond to the highest peak value of $G_{bi}^{(2)}(z)$ for each crystal thickness (see Fig. 6).

# VI. CONCLUSION

Focusing spontaneous parametric down-conversion (SPDC) through a thin lens yields a biphoton-excitation distribution that is the same as that of one-photon excitation at the pump wavelength in the axial and transverse directions. The two systems therefore have the same resolution when used in a scanning microscope. On the other hand, biphoton excitation in the image region of an ideal imaging optical system yields a distribution whose axial width is the crystal equivalent thickness $\ell_{eq} = \dfrac{\ell}{n_o(\omega_o)}$, and whose transverse width is equal to the transverse width of the pump at the input to the nonlinear crystal (NLC). When diffraction arising from the lens aperture is taken into account, the axial distribution widens. Including the broad spectrum of SPDC results in a narrowing of the axial distribution.

This work was supported by the National Science Foundation; by the Center for Subsurface Sensing and Imaging Systems (CenSSIS), an NSF engineering research center; and by the David & Lucile Packard Foundation.

## Captions:

**FIG. 1.** Two-photon excitation in the focal region using: (a) a two-photon source; (b) a biphoton source generated via SPDC in a nonlinear crystal (NLC) of thickness $\ell$. The spectral filter is narrow and centered about the degenerate wavelength $\lambda_o = 2\lambda_p$. $\theta_{SPDC}$ is 6/4/200110/9/2001the angle of the cone of SPDC emission.

**FIG. 2.** Comparison between distributions of one-photon, two-photon, and biphoton excitation. (a) Axial distribution (arbitrary units); the abscissa is normalized through division by $z_c$, where $z_c = 2\lambda_p F_\#^2$. (b) Transverse distribution (arbitrary units); the abscissa is normalized through division by $x_c$, where $x_c = 2\lambda_p F_\#$.

**FIG. 3.** Biphoton excitation in the imaging region.

**FIG. 4.** Axial distribution of biphoton excitation for the ideal imaging system, $G_{bi}^{(2)}(z)$, in arbitrary units. BBO crystals of thickness $\ell = 2$, 5, and 10 mm cut for collinear SPDC are considered. For a pump of wavelength $\lambda_p = 532$ nm, the collinear cut-angle is $\theta_{cut} = 22.88°$.

**FIG. 5.** The axial distribution of biphoton excitation for the ideal imaging system, $G_{bi}^{(2)}(z)$, in arbitrary units. The cut angle of a 2-mm BBO NLC is changed from collinear case to non-collinear case. The abscissa is normalized through division by $\ell_{eq}$.

**FIG. 6.** Dependence of the peak height of the axial biphoton excitation, $G_{bi}^{(2)}(z)$, on the cut-angle $\theta_{cut}$ in arbitrary units. Three BBO nonlinear crystals, with thicknesses $\ell = 2$, 5, and 10 mm are considered. The pump's wavelength is $\lambda_p = 532$ nm.

**FIG. 7.** Axial distribution of biphoton excitation $G_b^{(2)}(0,z)$, for different values of the ratio $\delta = \dfrac{\theta_{SPDC}}{\theta_{lens}}$ in arbitrary units. The parameter $\delta$ is the ratio between the angle of the cone of SPDC emission, $\theta_{SPDC}$, and the angle subtended by the lens, $\theta_{lens} = \dfrac{M}{2(M+1)F_\#}$. The abscissa is normalized through division by $\ell_{eq}$.

**FIG. 8.** Normalized axial distribution of biphoton excitation, $G_b^{(2)}(0,z)$, in arbitrary units for different biphoton bandwidths: (a) 10-mm BBO NLC cut at an angle $\theta_{cut} = 22.89°$. (b) 2-mm BBO NLC cut at an angle $\theta_{cut} = 22.95°$. For each plot the biphoton bandwidth value (BW) is normalized to the degenerate angular frequency $\omega_o = \dfrac{2\pi c}{\lambda_o}$, where $\lambda_o = 2\lambda_p = 1064$ nm is the degenerate wavelength. At this wavelength, the normalized frequency bandwidths 0.02, 0.06, 0.12, and 0.20 correspond to wavelength bandwidths 21, 64, 128, and 214 nm, respectively.

**FIG. 9.** Relative dependence of the peak of $G_b^{(2)}(0,z)$ on biphoton bandwidth, BW, for a 2-mm BBO NLC cut at an angle $\theta_{cut} = 22.95°$ (solid) and a 10-mm BBO NLC cut at an angle $\theta_{cut} = 22.89°$ (dotted). The BW is normalized to the degenerate angular frequency $\omega_o = \dfrac{2\pi c}{\lambda_o}$, where $\lambda_o = 2\lambda_p = 1064$ nm is the degenerate wavelength.

**Figures:**

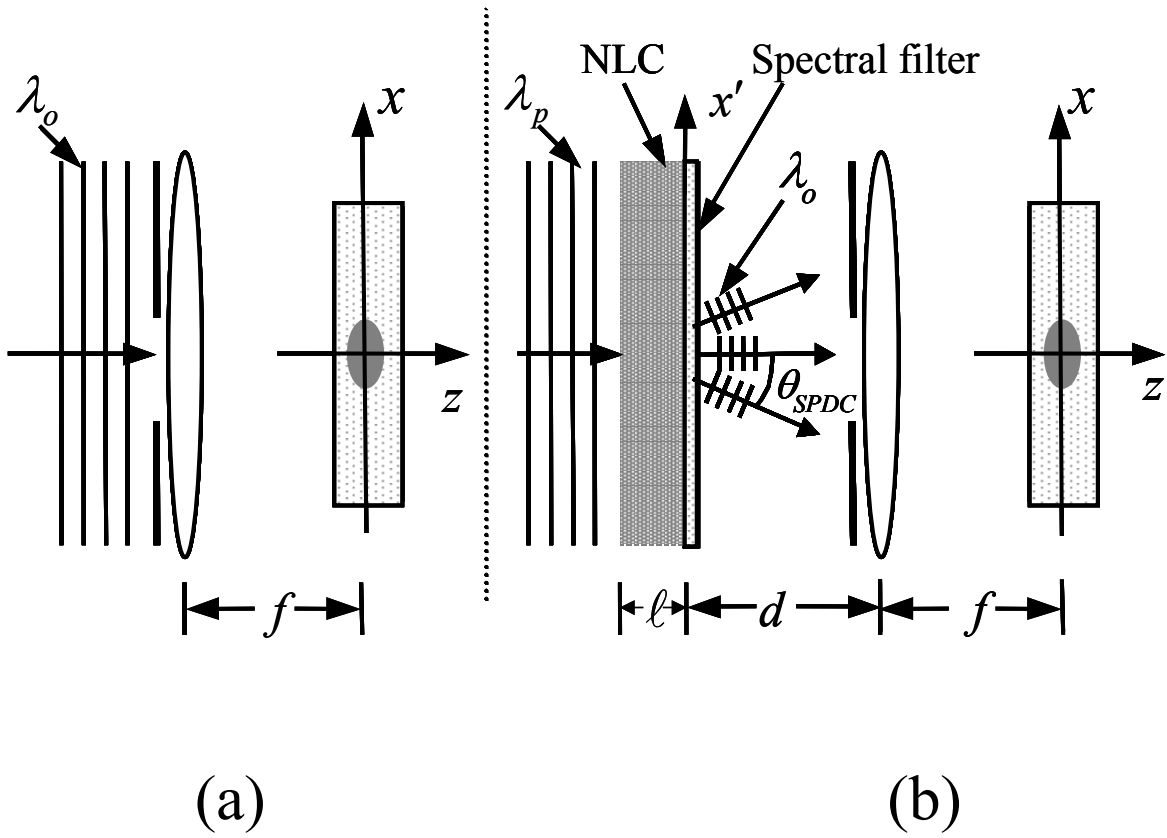

(a)                    (b)

Fig. 1

Figure 1, Magued B. Nasr

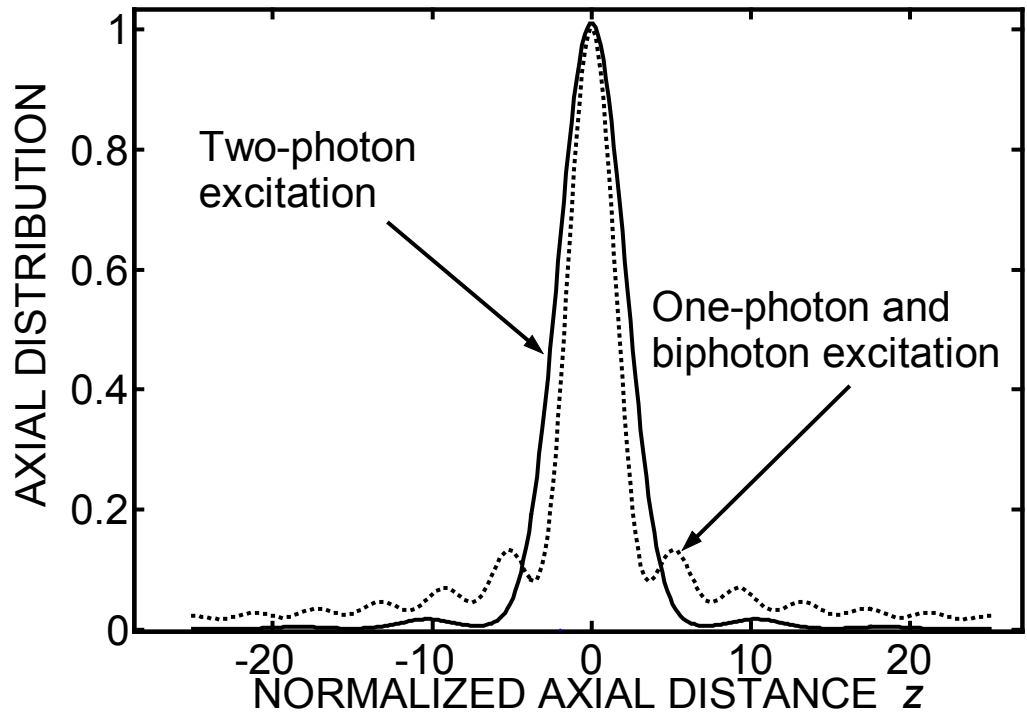

## (a)

## Fig. 2

Figure 2(a), Magued B. Nasr

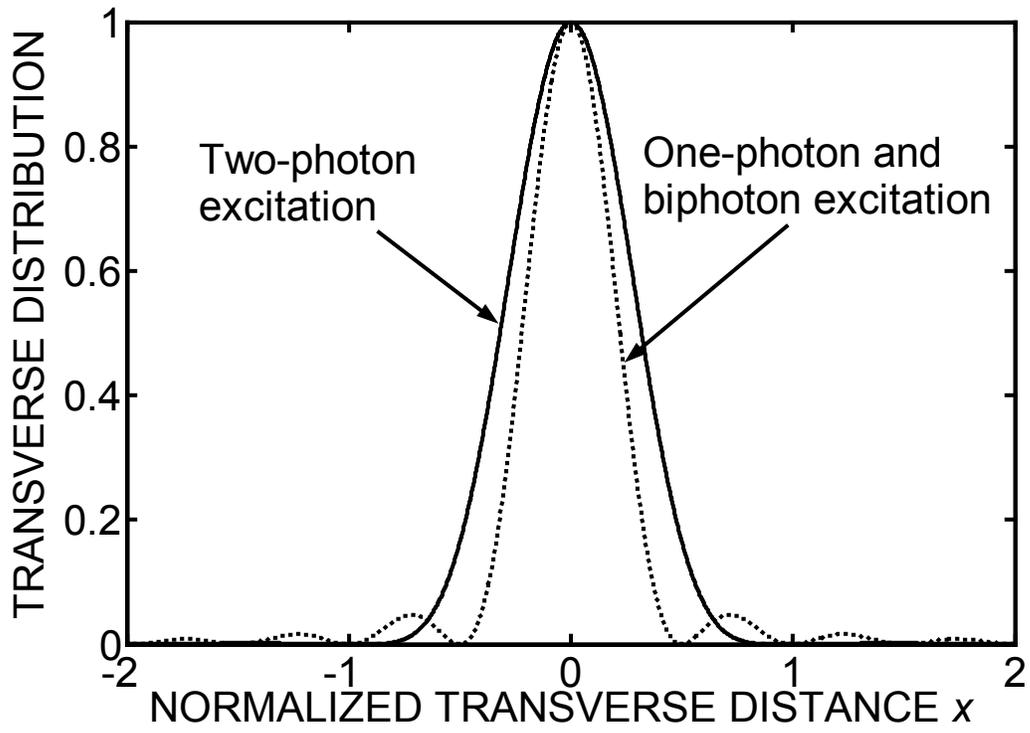

(b)

Fig. 2

Figure 2(b), Magued B. Nasr

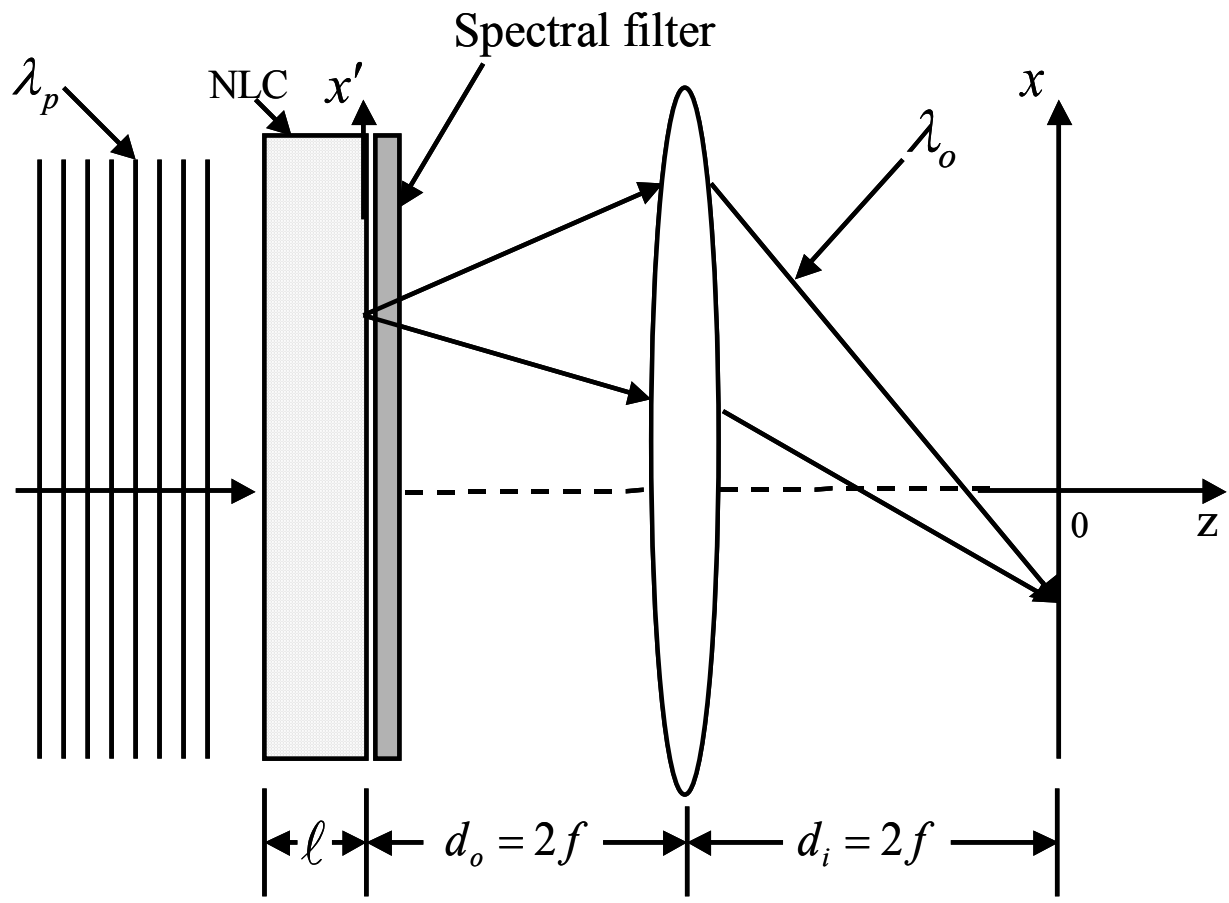

$\lambda_p$   NLC   $x'$   Spectral filter   $\lambda_o$

$x$

$0$   z

$\ell$   $d_o = 2f$   $d_i = 2f$

Fig. 3

Figure 3, Magued B. Nasr

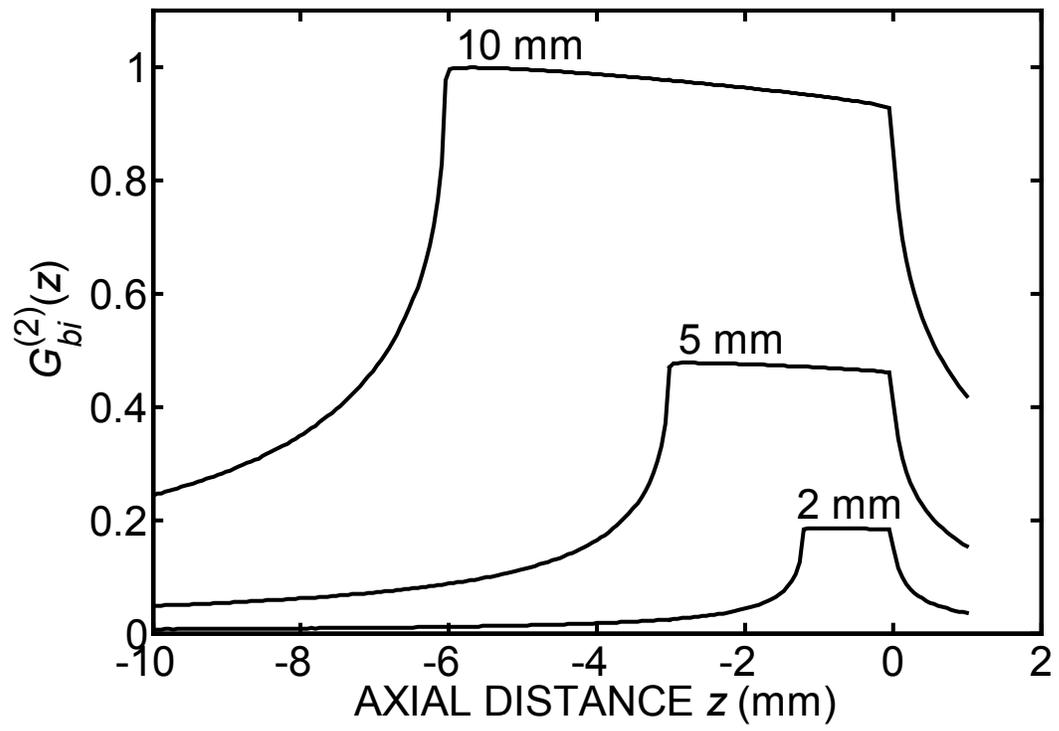

Fig. 4

Figure 4, Magued B. Nasr

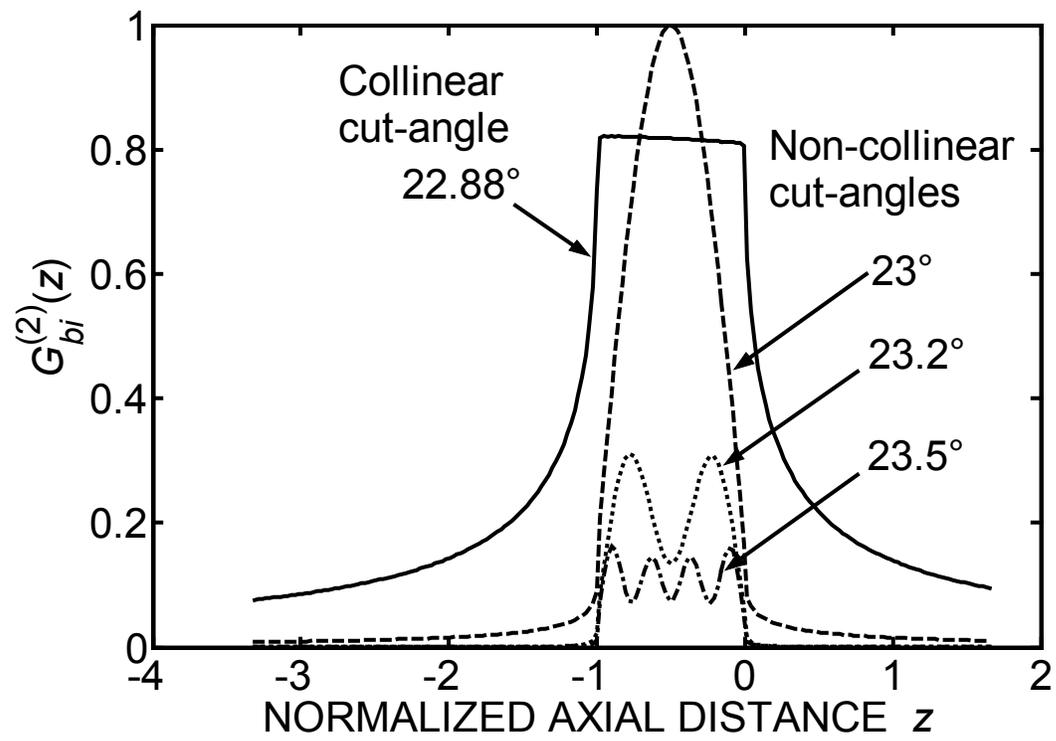

Fig. 5

Figure 5, Magued B. Nasr

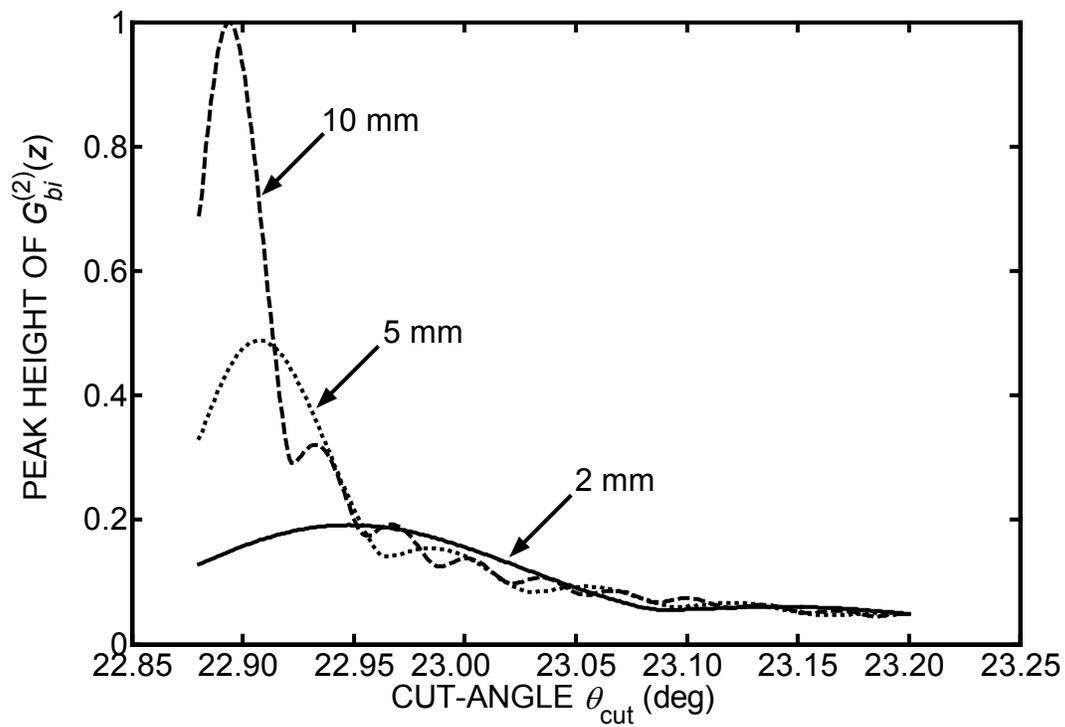

Fig. 6

Figure 6, Magued B. Nasr

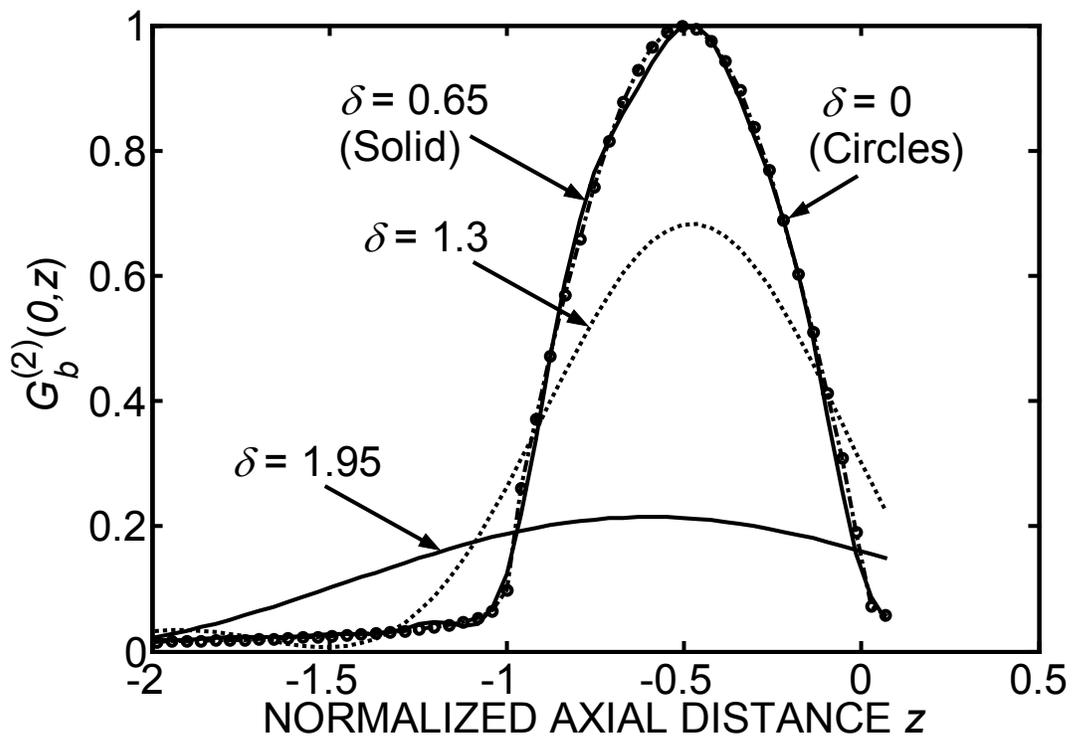

Fig. 7

Figure 7, Magued B. Nasr

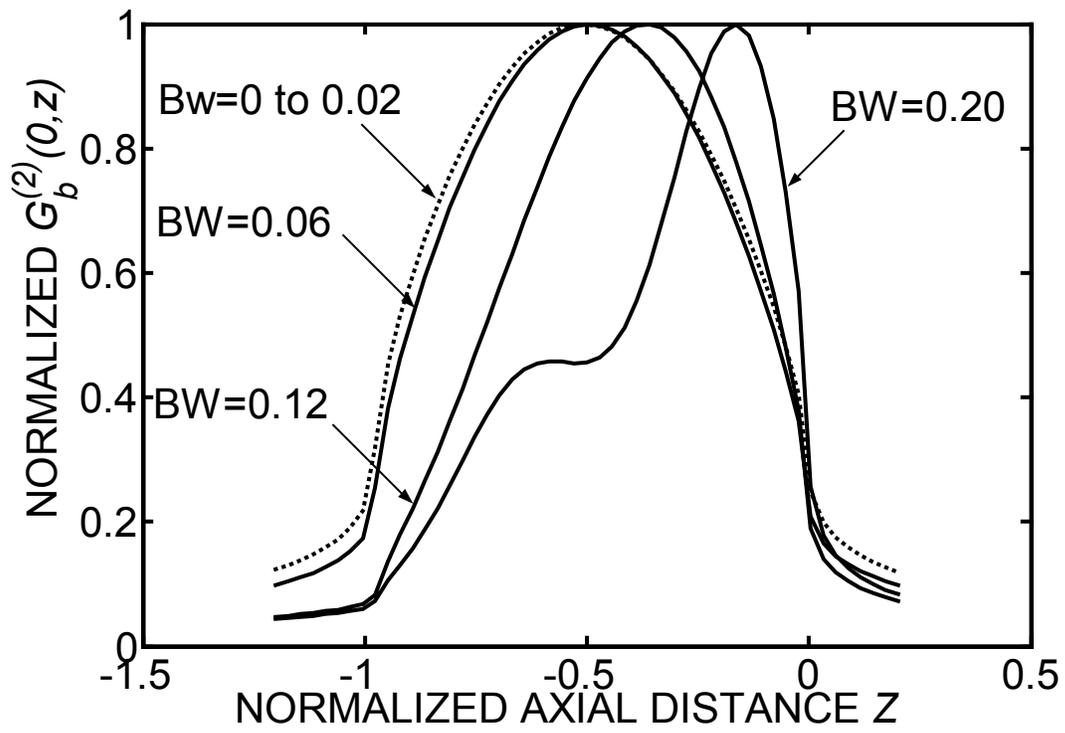

(a)

Fig.8

Figure 8(a), Magued B. Nasr

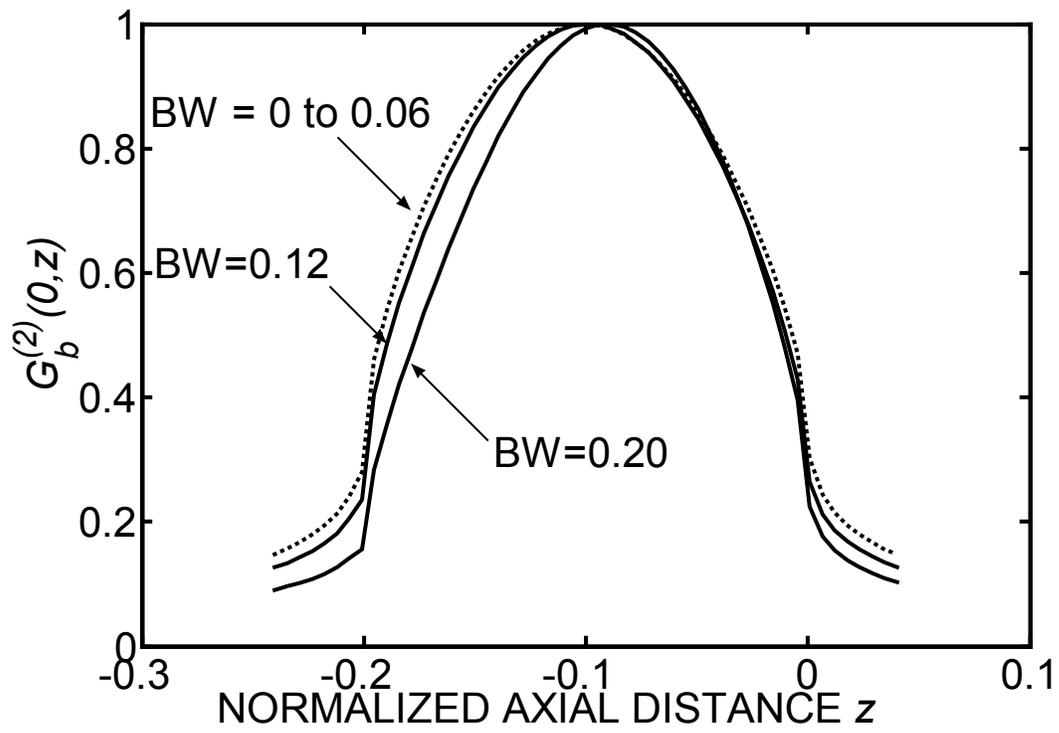

(b)

Fig. 8

Figure 8(b), Magued B. Nasr

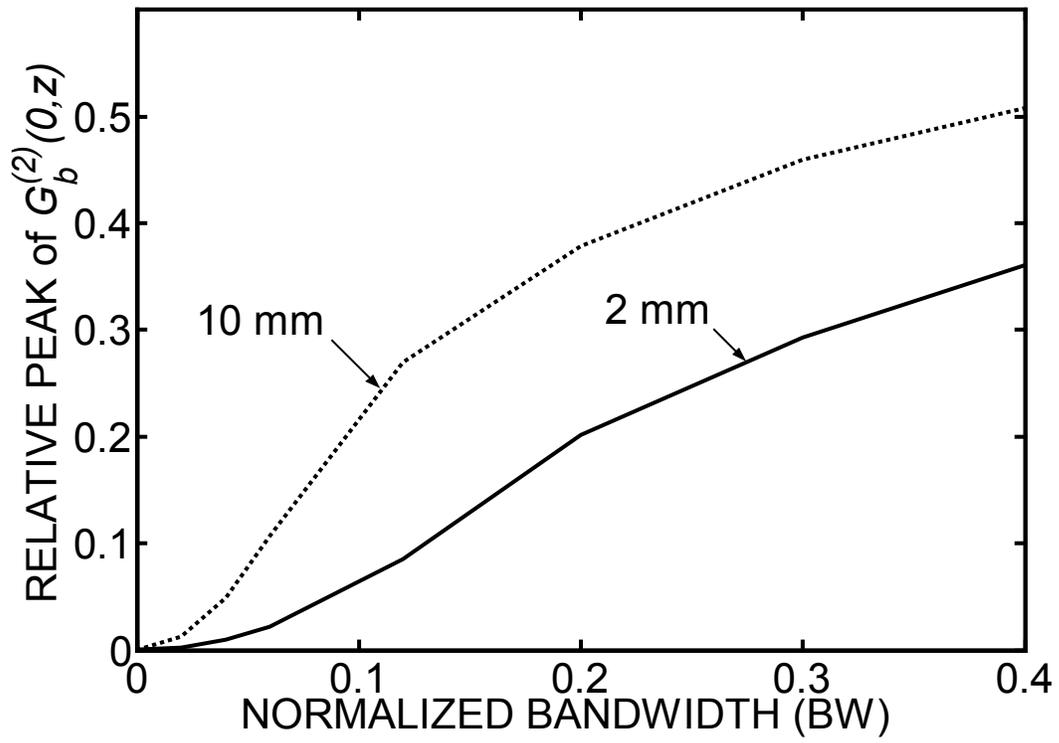

Fig. 9

Figure 9, Magued B. Nasr